\documentclass[a4paper,11pt]{article}
\usepackage{pos}
\usepackage{url}

\usepackage{natbib}
\usepackage{adjustbox}
\usepackage{subfig}
\usepackage{booktabs}

\title{Progress in generating gauge ensembles with Stabilized Wilson Fermions}

\author[a]{F.~Cuteri}
\author*[b]{A.~Francis}
\author[c]{P.~Fritzsch}
\author[d]{G.~Pederiva}
\author[e,f]{A.~Rago}
\author[g]{A.~Shindler}
\author[h]{A.~Walker-Loud}
\author[i]{S.~Zafeiropoulos}

\affiliation[a]{Institut für Theoretische Physik, Goethe Universität, 
Max-von-Laue-Str. 1, 60438 Frankfurt, Germany}

\affiliation[b]{Institute of Physics, National Yang Ming Chiao Tung University, 30010 Hsinchu, Taiwan}

\affiliation[c]{School of Mathematics, Trinity College Dublin, Dublin 2, Ireland}

\affiliation[d]{J\"ulich Supercomputing Centre, Forschungszentrum J\"ulich GmbH, 52428 J\"ulich, Germany}

\affiliation[e]{IMADA \& QTC, University of Southern Denmark, Odense, Denmark}
\affiliation[f]{Theoretical Physics Department, CERN, CH-1211 Geneva 23, Switzerland}

\affiliation[g]{TTK, RWTH Aachen University, 52056 Aachen, Germany}

\affiliation[h]{Nuclear Science Division, Lawrence Berkeley National Laboratory, Berkeley, CA 94720, USA}

\affiliation[i]{Aix Marseille Univ, Universit\'e de Toulon, CNRS, CPT, Marseille, France}

\abstract{The continued generation of $N_f=2+1$ quark flavor gauge configurations using stabilized Wilson fermions by the open lattice initiative (OpenLat) is reported. We present the status of our ongoing production and show updates on increasing statistics at the four lattice spacings $a=0.12, 0.094, 0.077$ and $0.064$ fm. Aside from the $SU(3)$ flavor symmetric point we discuss advancements in going towards physical pion masses. We show preliminary results of the pion decay constants, extending previous results, and discuss further validation observables on the available ensembles.
}

\FullConference{The 40th International Symposium on Lattice Field Theory (Lattice 2023)\\
July 31st - August 4th, 2023\\
Fermi National Accelerator Laboratory\\}

\begin{document}
\maketitle

\section{Introduction and overview}

In this proceedings article we report on the progress of the open lattice initiative (OpenLat)\footnote{https://openlat1.gitlab.io} in producing gauge ensembles at a large scale and for broad use in the community. We proceed with our outlined plan \cite{Cuteri:2022oms} to generate and, in the near future, provide gauge ensembles with $N_f=2+1$ flavors of Wilson fermions. All our calculations are performed in the stabilized Wilson fermion (SWF) framework \cite{Francis:2019muy}: That is, we use and improved gauge action, for example the L\"uscher-Weisz improved gauge action \cite{Weisz:1982zw,Luscher:1984xn,Curci:1983an} in our case, and the exponentiated Clover action \cite{Francis:2019muy} for the quarks with non-perturbatively tuned clover coefficient \cite{Francis:2019muy,Francis:2022hyr}. Gauge configurations are generated by means of the stochastic molecular dynamics (SMD) algorithm \cite{Horowitz:1985kd,Horowitz:1986dt,HOROWITZ1991247,Jansen:1995gz} with quad precision sums and the supremum norm as stopping criterion. 

With this numerical set-up we choose the lattice parameters such that we cover a broad window of lattice spacings $a=0.12~$fm to $a=0.064~$fm\footnote{A further lattice spacing of $a=0.055~$fm is part of our production plan and is currently also being tuned. However, no results will be shown here.} and quark mass parameters corresponding to pion masses of $m_\pi\leq 412~$MeV to $m_\pi \sim 135~$MeV, while $m_\pi L \gtrsim 4$ and $L\gtrsim 3~$fm. Further quality criteria and test observables for the stability and health of the simulations have been laid out \cite{Francis:2022hyr} and are continuously being refined \cite{Cuteri:2022oms}.
The scale is set via the gradient flow and converted to physical units using the value $\sqrt{8t_0} = 0.414(5)~$fm \cite{Bruno:2016plf}. The mass parameters are chosen along a chiral trajectory where Tr$[M]$=constant, i.e. the sum of bare quark masses is held fixed after having been tuned to a reference physics point. This is the so-called $SU(3)$-flavor symmetric point ($SU(3)_F$) where $m_\pi=m_K=412~$MeV. For further details see \cite{Bietenholz:2010jr,Bruno:2014jqa,Strassberger:2021tsu,Francis:2022hyr}.

\begin{figure}
\centering
\includegraphics[width=0.4\textwidth]{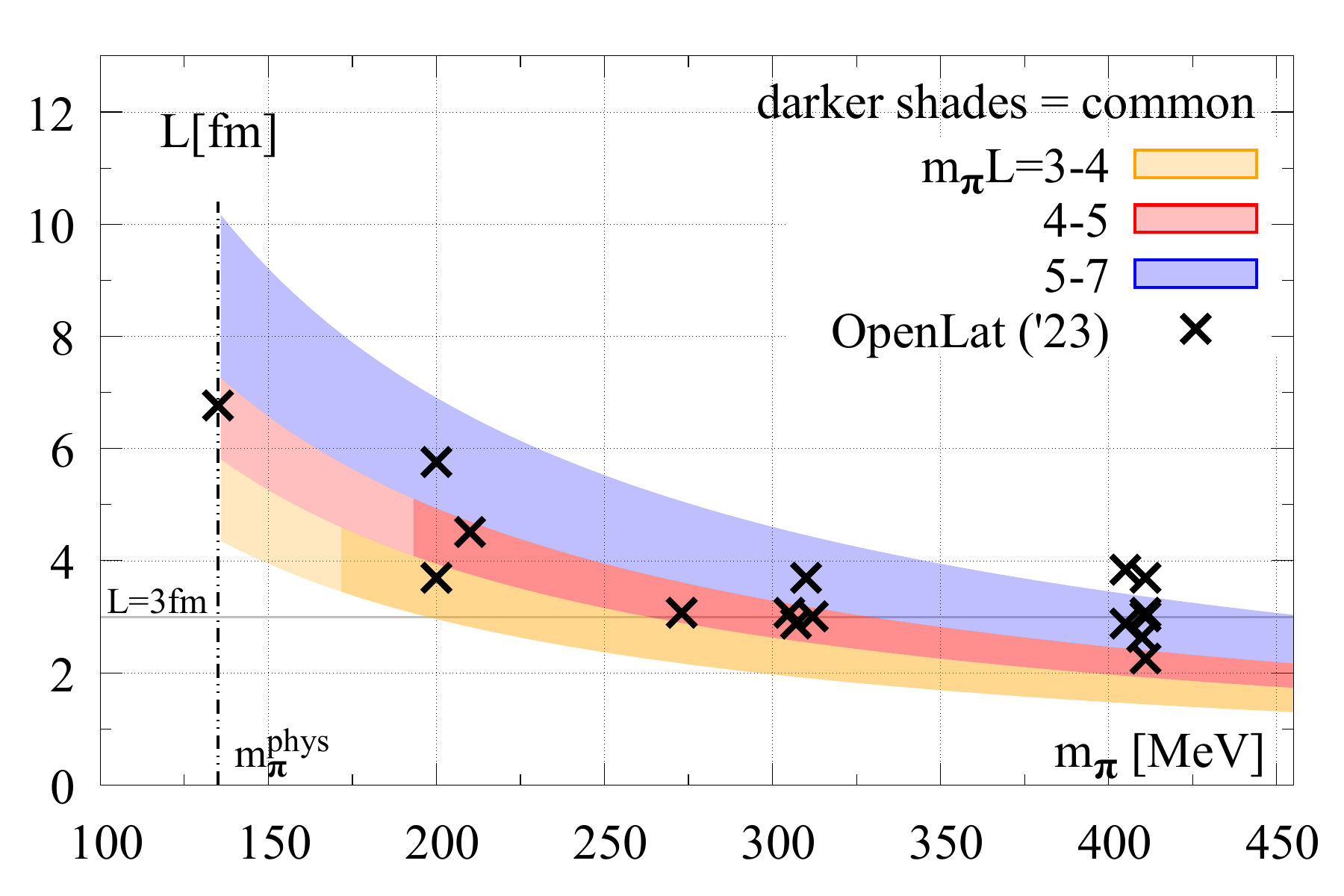}
\includegraphics[width=0.43\textwidth]{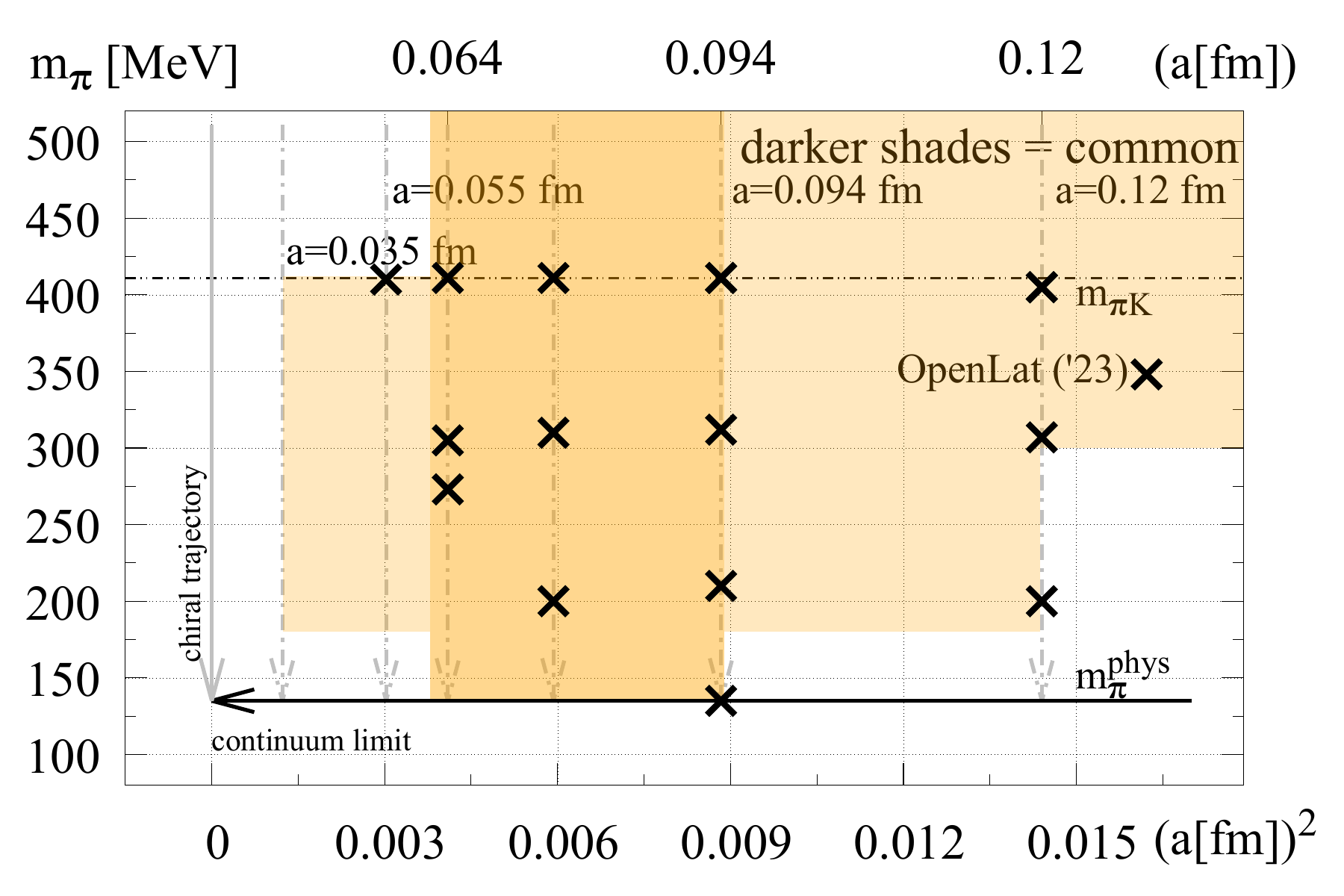}
\caption{Overview of OpenLat lattice ensembles currently available or in production. Left: Ensembles shown in terms of volume over pion mass. Right: In terms of pion mass over lattice spacing. See text for further explanation. }
\label{fig:tmp1}
\end{figure}

An overview of the parameter window mapped out by OpenLat is shown in Fig.~\ref{fig:tmp1}. In the left panel we show our ensembles, available and under production, in terms of the pion mass and the volume, where the colored bands denote the regions of $m_\pi L$ their combination implies. In the right panel we show this information instead in terms of lattice spacing and the pion mass.
The darker shades provide an indication of the most common regions covered in the broader lattice community, the lighter shades denote regions that we are hoping or are currently expanding to.

In this article we sidestep further explanations of the lattice details and refer to \cite{Francis:2019muy,Francis:2022hyr,Bali:2022mlg,Cuteri:2022oms} for more in-depth discussion. The main updates reported on here are: Firstly, an update on the generation status and where we are in our production plan. Secondly, we present first results on evaluating the stability against changes of the sign of the reweighting factors for strange and light quarks in our choice of algorithmic set-up used for the production of gauge configurations. 
Next, we give an update on reaching physical point simulations with $m_\pi=m_{\pi, \text{phys}}=135~$MeV at $a=0.094~$fm and $m_\pi L\gtrsim 4$.
Finally, we present the first determination of the renormalized pion decay constant $F_\pi$ at the $SU(3)_F$ point on all available lattice spacings. This updates a previous determination at $a=0.094~$fm and $0.064~$fm and significantly increases the precision reached, in addition to adding new results at the lattice spacings $a=0.12~$fm and $a=0.077~$fm. 
All shown results and numbers are preliminary.

\begin{table}[t!]
\centering
\begin{tabular}{lcclllll}
\hline
\hline
Label & Lattice Size & $m_\pi~$[MeV] & $m_\pi L$  & $a~$[fm] & $L~$[fm] & Statistics  \\ \midrule
\texttt{a12\_m412\_mL6.0} & $96\times 24^3$ & 412 & 6.01 & 0.12 & 2.88 & 1500 \\ 
\texttt{a12\_m300\_mL4.5} & $96\times 24^3$ & 307 & 4.46 & 0.12 &  2.88& 800  \\
\midrule
\texttt{a094\_m412\_mL6.3} & $96\times 32^3$ & 410 & 6.25 & 0.094 &  3.01 & 1500  \\ 
\texttt{a094\_m300\_mL4.7} & $96\times 32^3$ & 310 & 4.73 & 0.094 & 3.01 & 500  \\ 
\midrule
\texttt{a077\_m412\_mL7.7} & $96\times 48^3$ & 411 & 7.70 & 0.077 &3.70 & 1500  \\ 
\texttt{a077\_m300\_mL5.8} & $96\times 48^3$ & 307  & 5.75 & 0.077 &3.70 & 400  \\ \midrule
\texttt{a064\_m412\_mL6.4} & $96\times 48^3$ & 411  & 6.40 & 0.064 &3.07 & 1000  \\
\texttt{a064\_m300\_mL4.8} & $96\times 48^3$ & 307  & 4.78 & 0.064 &3.07& 800  \\
 \hline
\hline
\end{tabular}
\caption{ Listed are all ensembles at production level and their statistics in units of $N= N_{\text{MDU}} / \tau_Q$, where $N_{\text{MDU}}$ is the total number updates performed and $\tau_Q$ is the integrated autocorrelation time of the topological charge rounded up to its nearest integer, both in MDU units.}
\label{tab:ensembles}
\end{table}

\section{Production update}

Our generation plan can be summarised in three stages \cite{Francis:2022hyr}:
\begin{itemize}
\itemsep0ex
\item [{\bf 1.}] {\bf Stage:} Perform high precision tuning and generate ensembles with 3 dynamical quarks at the flavor symmetric point with high statistics ($N=500+$, often 1000+) at the lattice spacings $a=0.12, 0.094, 0.077$ and $0.064$~fm.
\item [{\bf 2.}] {\bf Stage:} Reduce the light quark masses with Tr$[M]$=constant. The goal is to have matched pion passes at $m_\pi\simeq 300, 200$~MeV for $a=0.12, 0.094, 0.077$ and $0.064$~fm. Add the lattice spacing $a=0.055~$fm (open boundary conditions \cite{Luscher:2011kk}) at this stage.
\item [{\bf 3.}] {\bf Stage:} Go towards the physical values of the pion mass $m_\pi=135$~MeV on at least one lattice spacing, with extension to all lattice spacings in the long term.
\end{itemize}

Part of the OpenLat goals is to generate state-of-the-art QCD gauge ensembles for physics applications and share them with the community, compatible with the ILDG standard, see e.g. \cite{Karsch:2022tqw,Bali:2022mlg} and https://hpc.desy.de/ildg/, without embargo time after publication, or before publication on a case-by-case basis. As such, an important component of our effort is for the data to adhere to an open science policy and in particular the FAIR principles, i.e. the data are to be made Findable, Accessible, Interoperable and Reusable (https://www.go-fair.org/). 

To each generation process there are two separate levels: The tuning and the production level.
In the tuning level thermalization is performed and the algorithm's parameters are adjusted over a large number of SMD updates. Within the SMD algorithm there is a direct relation of the update cycles of the algorithm to the more traditionally used molecular dynamics units (MDU). 
We estimate the autocorrelation time in the topological charge $\tau_Q$ in MDU units and round it up to the nearest integer. We then gather $N_{\rm tune}=100\cdot \tau_Q$ updates in units of MDU for further testing.
From that point onwards, and with the run being stable, we consider this a production level ensemble and continue generating configurations in the tuned setting.

Within the production plan and strategy we are currently completing stage 1, with the publication pending, and are in the process of finishing up the production of the first stage 2 ensembles at $m_\pi\simeq 300~$MeV. An overview of the production ensembles is given in Tab.~\ref{tab:ensembles}. 

\begin{figure}
\centering
\includegraphics[width=0.34\textwidth]{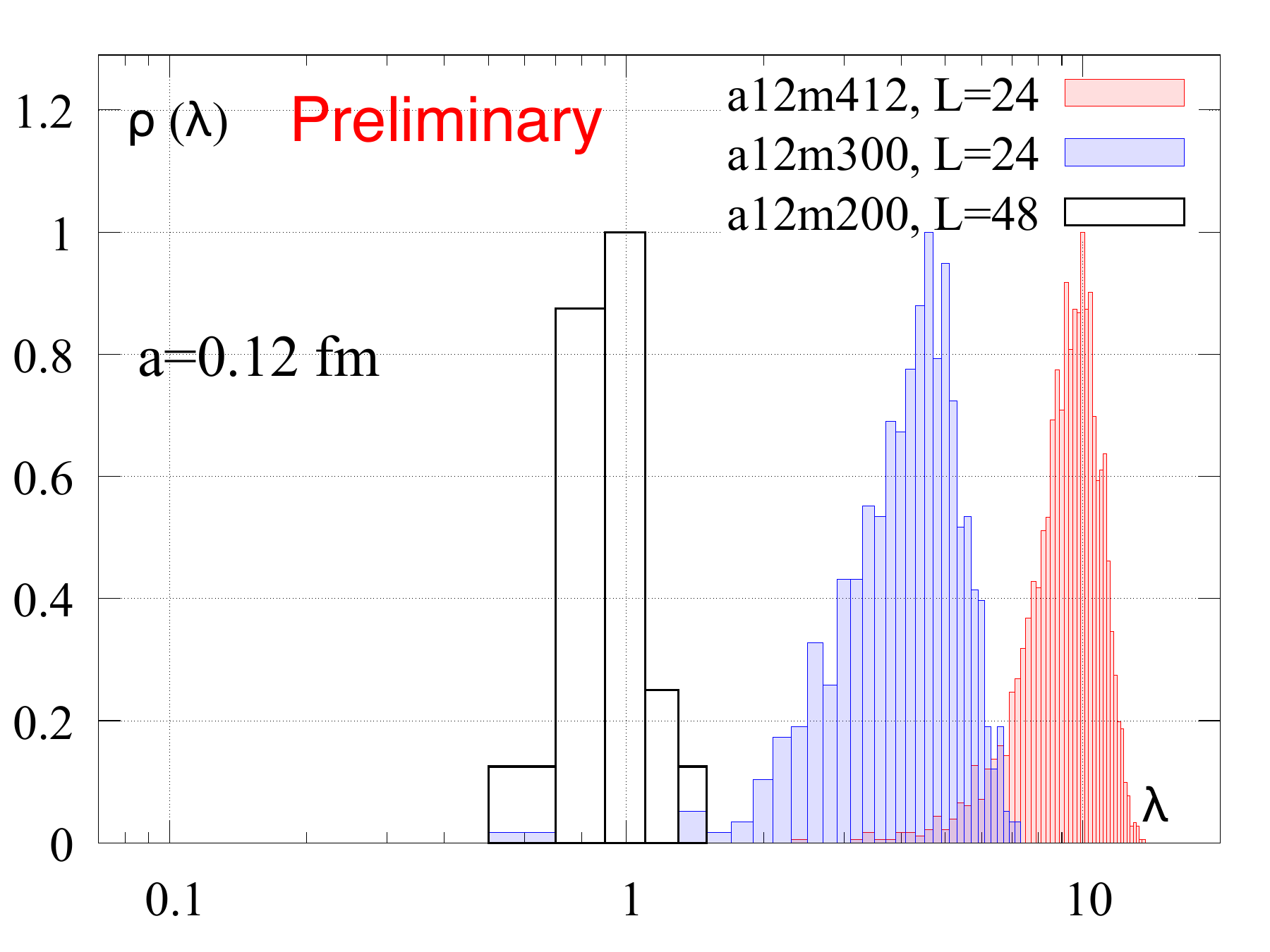}
\includegraphics[width=0.34\textwidth]{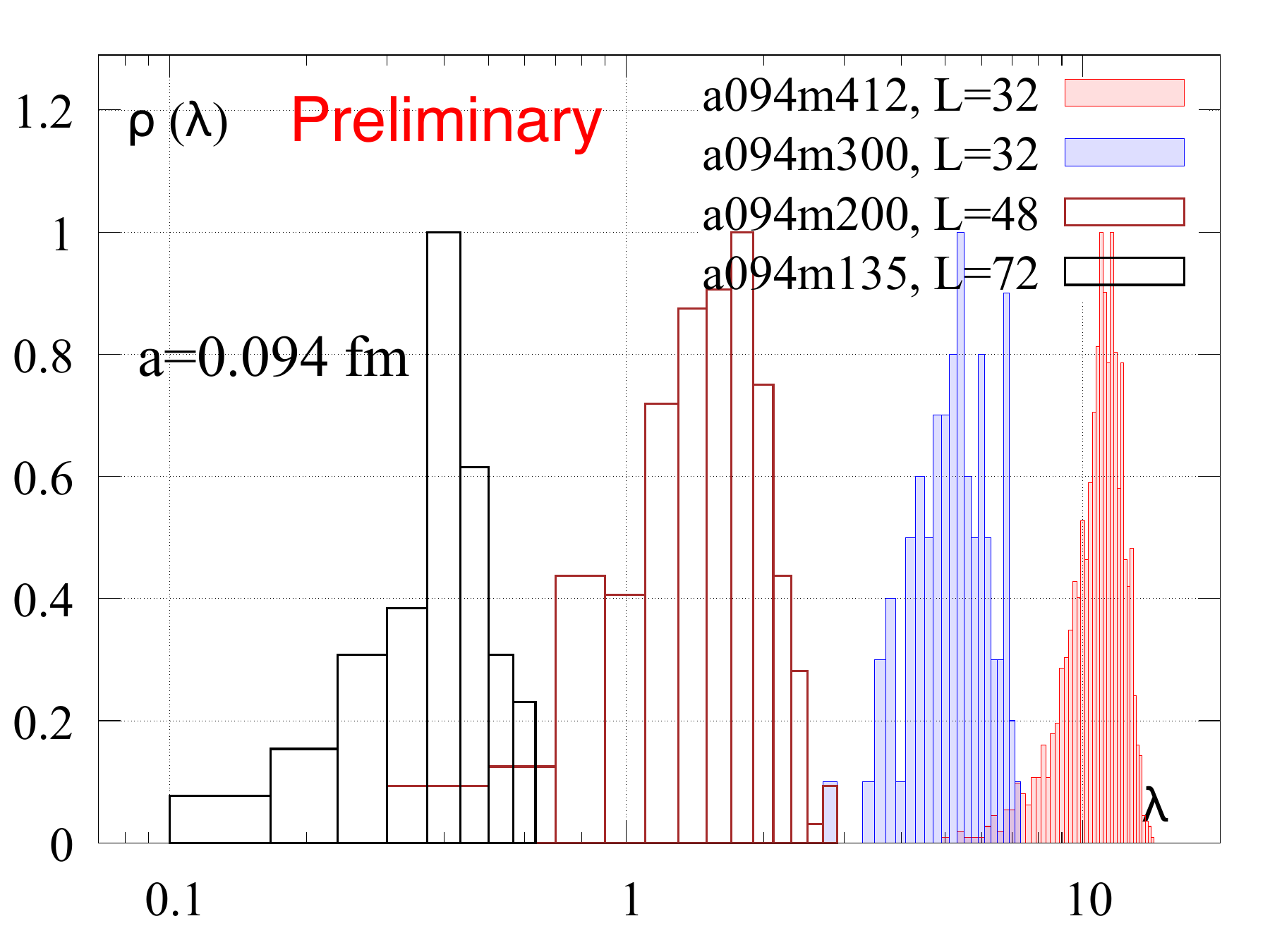}\\
\includegraphics[width=0.34\textwidth]{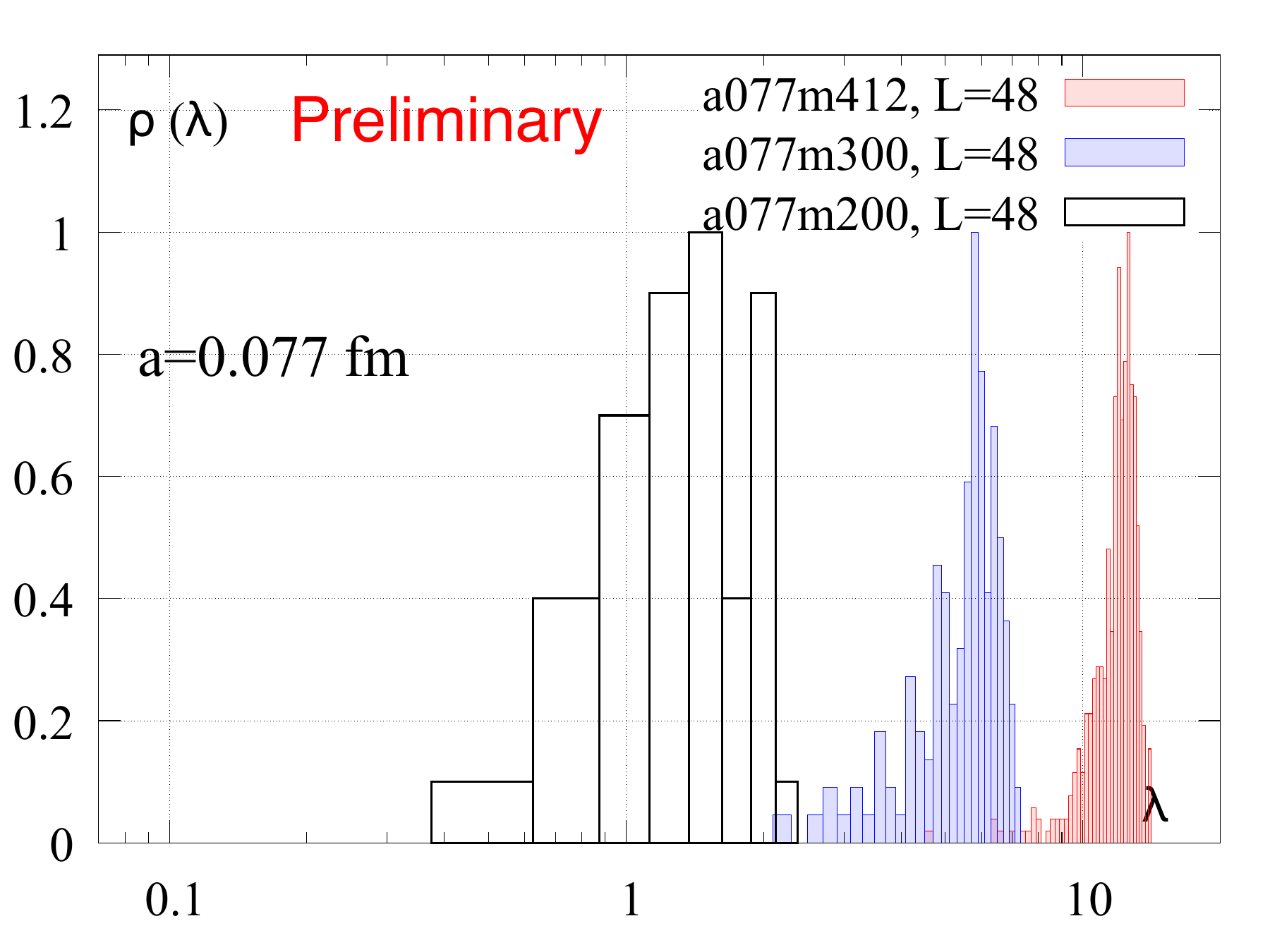}
\includegraphics[width=0.34\textwidth]{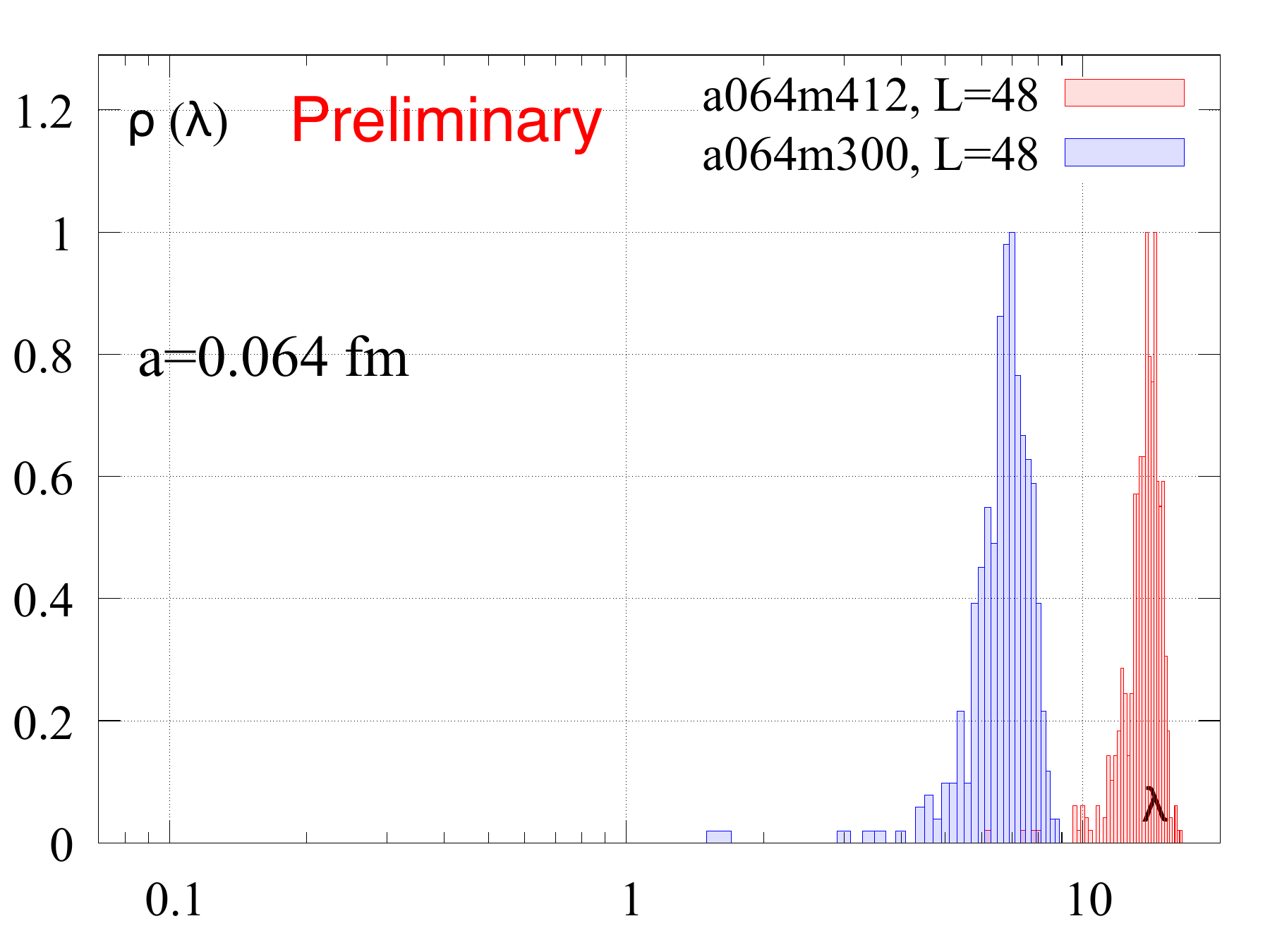}
\includegraphics[width=0.49\textwidth]{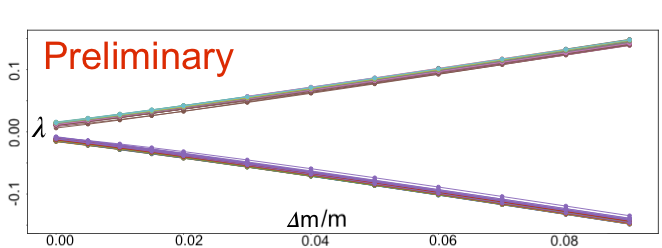}
\includegraphics[width=0.49\textwidth]{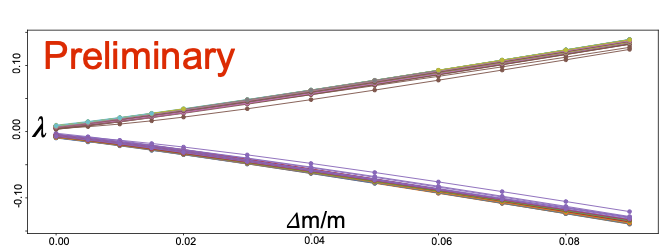}
\caption{Top two rows: Distributions $\rho(\lambda)$ of lowest eigenvalues of the lattice Dirac operator $\lambda(\sqrt{D^\dagger D})$. Bottom row: Eigenvalue pairs of the Hermitian $\lambda(\hat Q)$ over the normalised valence quark mass splitting on a subset of the \texttt{a094\_m412} (left) and \texttt{a094\_m300} (right) ensembles.}
\label{fig:tmp2}
\end{figure}

To further illustrate the production levels, including some tuning stage ensembles, we show the distribution of the lowest eigenvalue of the lattice Dirac operator $\lambda(\sqrt{D^\dagger D})$ in Fig.~\ref{fig:tmp2} (top 2 rows). Each panel shows the results from a different lattice spacing, with the different colors denoting the pion masses available. Note that the distributions have been normalised to their maxima for legibility. The main stage 1 and 2 updates are given in red and blue, these are the production level ensembles. The results for three more ensembles at the tuning stage are given to further highlight our progress: \texttt{a12\_m200} and \texttt{a094\_m200}, which will not be further mentioned, and \texttt{a094\_m135}, for which we include more details below.
In all cases we note that we are observing a well peaked distribution, with a visible mass gap and no near-zero results.

\section{Regularity of the reweighting factors}

Wilson-type fermions by construction have a complicated relationship with chiral symmetry and recently in \cite{Mohler:2020txx} a further complication, hitherto not considered problematic, was highlighted. As such, in principle positivity of the fermion determinant is guaranteed by chiral symmetry and $\gamma_5$ hermiticity of the fermion operator for each flavor. However, in Wilson-type fermions this is not the case due to explicit chiral symmetry breaking and negative eigenvalues ($\lambda(\hat D)$) are possible. Approaching the continuum limit one can argue that configurations with such problems are very unlikely to be visited by standard update algorithms.
This situation is further complicated, however, in $N_f=2+1$ or $N_f=2+1+1$ simulations where the extra single flavors for the strange and charm quarks are implemented using the RHMC \cite{Clark:2006fx} or PHMC \cite{Frezzotti:1998eu,Frezzotti:1998yp} algorithms. Here, positivity of the fermion determinant is generally assumed due to the larger masses of the strange and charm quarks.
A violation of this assumption becomes visible through a negative sign of the reweighting factors that need to be computed for these algorithms, and it was precisely this 
which the authors of \cite{Mohler:2020txx} found in a sub-set of large scale calculations using Wilson-Clover fermions. This presents a problem since in the stochastic determination of reweighting factors the sign is not determined. 

To diagnose the problem a direct evaluation via the eigenvalues of the full lattice Dirac operator $\lambda(\hat D)$ would be desirable but is unpractical. An indirect but more practical test was suggested in \cite{Mohler:2020txx} by studying instead the eigenvalues of the Hermitian $\hat Q=\gamma_5 \hat D$.
In this setting eigenvalues appear in positive and negative pairs. Their splitting furthermore depends on the input (valence) quark mass chosen. A violation of the positivity of the fermion determinant then becomes visible as a mismatch in the pairs $\lambda(\hat Q, m_{\text{valence}})$. Indeed, as $m_{\text{valence}}$ increases we can expect to observe a zero-crossing in this case. We refer to \cite{Mohler:2020txx} for more details on the procedure and what to expect when a mismatch occurs.

\begin{figure}
\centering
\includegraphics[width=0.49\textwidth]{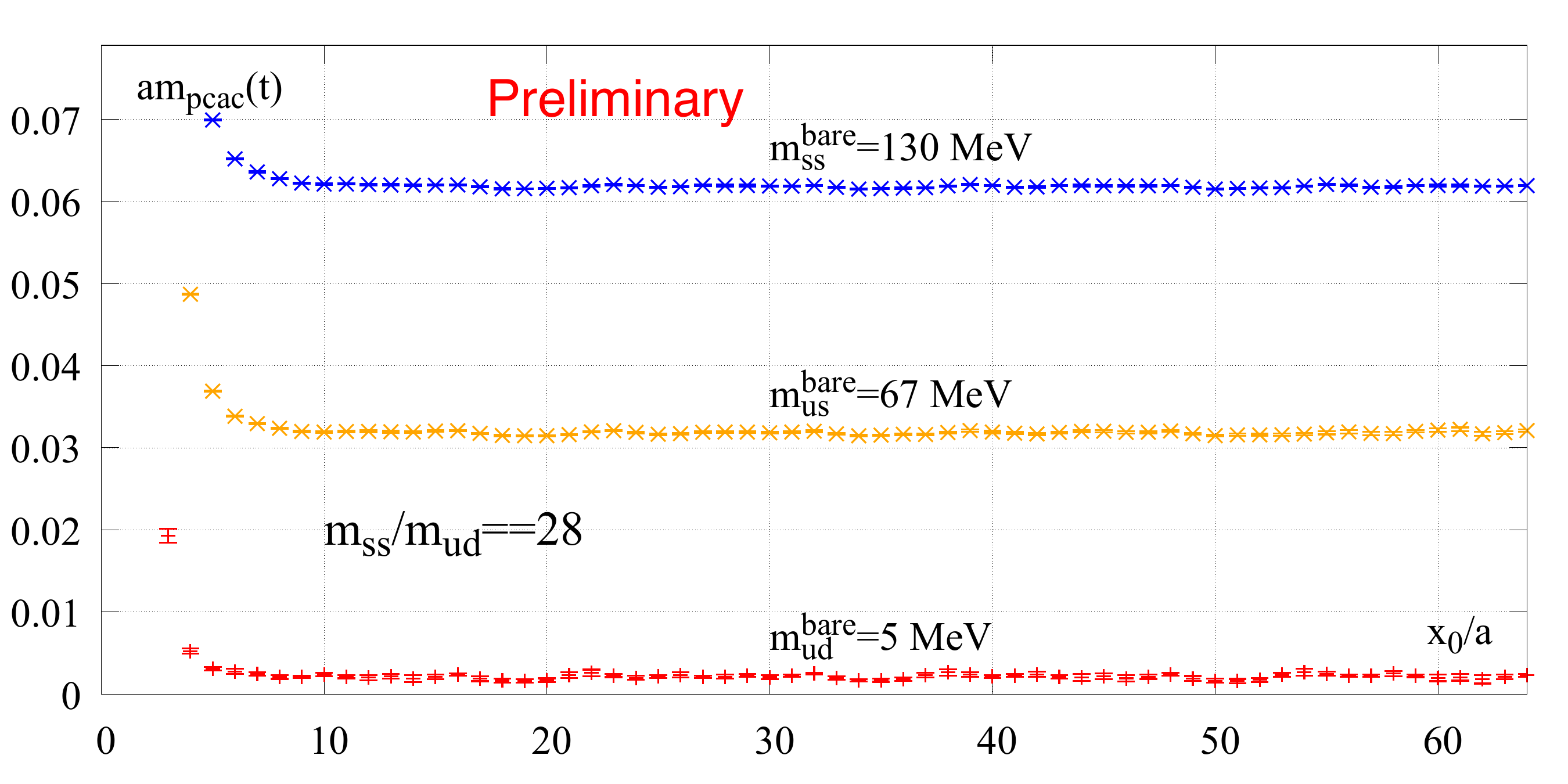}
\includegraphics[width=0.49\textwidth]{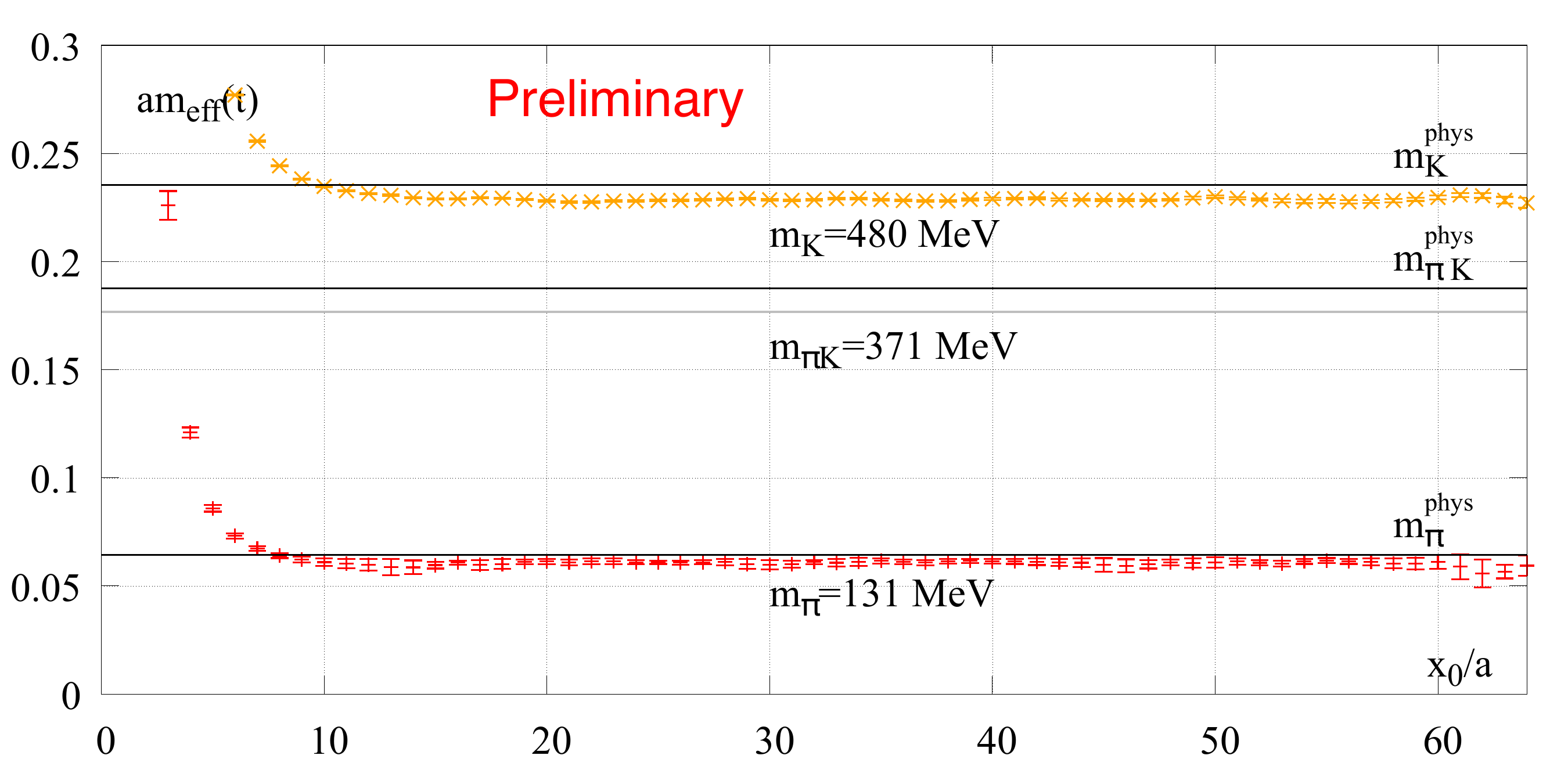}
\caption{First physical point results with $m_\pi L=4.6$ and $a=0.094~$fm. Left: $PCAC$ quark masses. Right: Pseudoscalar meson masses.}
\label{fig:tmp3}
\end{figure}

Results on the valence mass dependence of this quantity $\lambda(\hat Q)$ for the \texttt{a094\_m412} (left panel) and \texttt{a094\_m300} (right panel) are shown in Fig.~\ref{fig:tmp3} (bottom row), whereby in the latter we consider the light quark case. The horizontal axis hereby is given in terms of the splitting between the valence and the sea quark masses of the corresponding pion correlators over the sea quark value.
Our results are regular and we do not observe would-be negative signs on the set of configurations evaluated. 
It should be noted that this is not yet the full set and we are working to perform the analysis for all available ensembles and configurations. 
The results shown serve to indicate that we are now ready to process the full set of ensembles in the near future.


\section{Towards the physical point}

Tuning and production of ensembles proceed in parallel and while we are generating the \texttt{m412} and \texttt{m300} configurations we simultaneously focus on pushing towards the physical point at $m_\pi=135~$MeV. Here we report first success in this direction at $a=0.094~$fm with $m_\pi L=4.6$. 
To this extent we show results for the bare $m_{PCAC}$ quark masses (right) and the pseudoscalar mesons (left). In both cases $U(1)$ noise sources on a sparsened grid were used to evaluate the required correlation functions. With a small statistics of 20 independent configurations the results indicate we have reached a pion mass value of $m_\pi=131~$MeV. The corresponding kaon mass is $m_K=480~$MeV, which highlights one of the drawbacks of the strategy to use a fixed trace of the mass matrix, as corrections can lead to missing the physical mass kaon point. In the future we plan to correct this mismatch. Furthermore we will continue our program to gather 100 tuning configurations and, pending them passing all tests, proceeding to the production level.

\section{Statistics update on the calculation of $F_\pi$}

\begin{figure}
\centering
\includegraphics[width=0.45\textwidth]{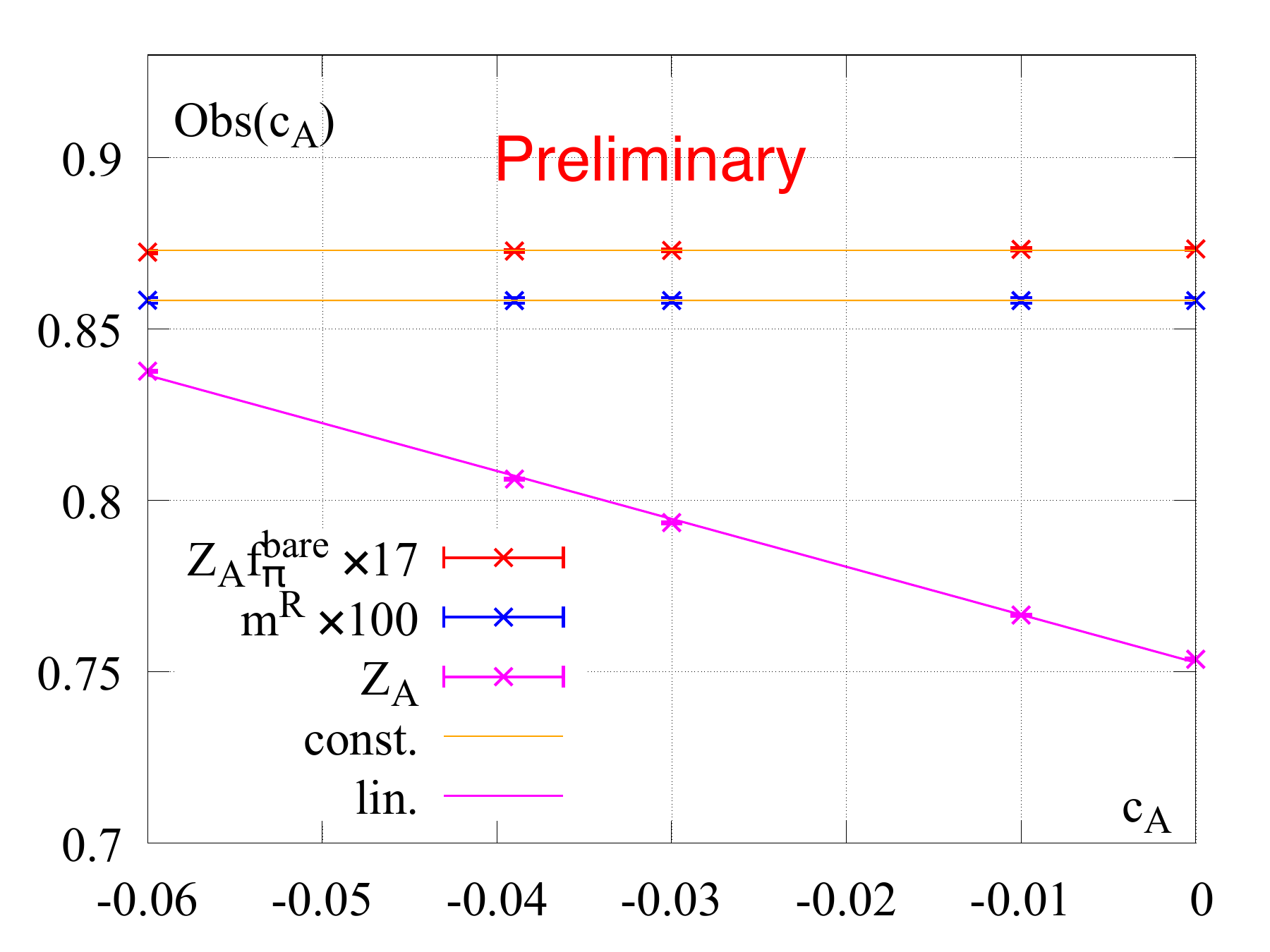}
\includegraphics[width=0.45\textwidth]{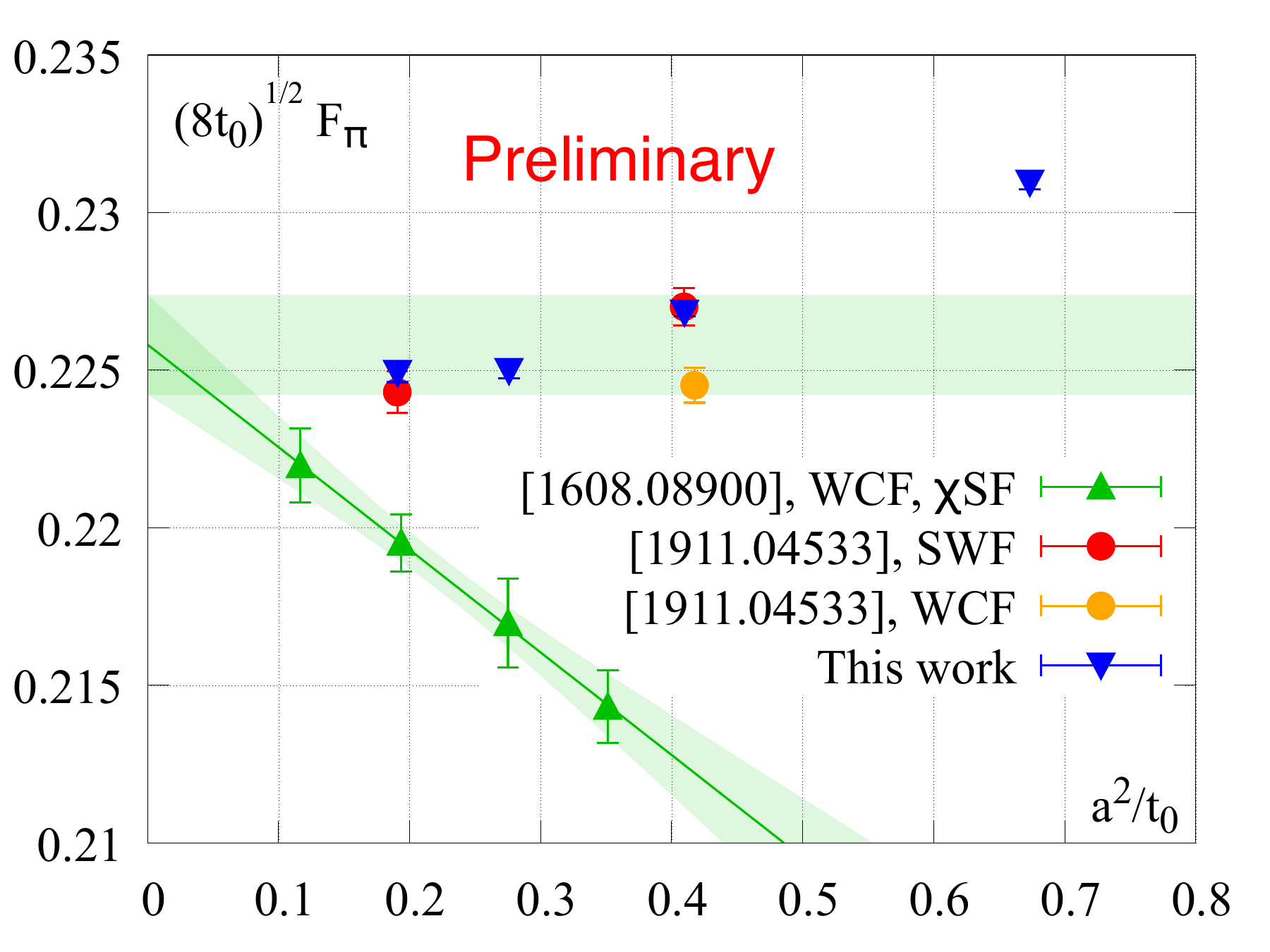}
\caption{Determining decay constants, quark masses and renormalization constants from the procedure using gradient flowed correlators. Left: Independence of $Z_A\cdot f_\pi^{\text{bare}}$ and $m^R$, as well as dependence of $Z_A$ on $c_A$. All results are  shown in lattice units and have been rescaled for legibility. Right: Lattice spacing dependence of $F_\pi$ from multiple calculations, see text.}
\label{fig:tmp4}
\end{figure}

The final update presented here concerns the calculation of the renormalized pion decay constant. Based on \cite{Luscher:2013cpa}, in \cite{Francis:2019muy} the calculation was performed by evaluating correlation functions with gradient flow. The basic idea is to determine the required renormalization factors by probing chiral symmetry at positive flow times through appropriate combinations of flowed and unflowed correlators. As outlined in \cite{Francis:2019muy} the renormalized decay constants in this set-up actually are insensitive to the improvement coefficient $c_A$, which is usually required as input from a separate calculation. 
Continuing to refer to the publication \cite{Francis:2019muy}, the key outcome is that while the renormalized decay constant $F_\pi$ and also the renormalized quark mass $m^R=Z_A \cdot m_{PCAC}^{\text{bare}}$ become insensitive to $c_A$, the value of $Z_A$ is not insensitive and does depend on it. As such we cannot quote its value without a separate determination of this improvement coefficient.
In Fig.~\ref{fig:tmp4} (left) we show the results for the three quantities $Z_A\cdot f_\pi^{\text{bare}}$, $m^R$ and $Z_A$ with their dependence on $c_A$, whereby all are shown in lattice units and have been rescaled for legibility. The former two quantities show negligible dependence, as expected. In Fig.~\ref{fig:tmp4} (right) we show preliminary results for $F_\pi$ at all available lattice spacings at the $SU(3)_F$ point from this calculation (blue). We confirm and update the precision on our previous results (red). In green we show results obtained using the traditional method on Wilson-Clover fermion ensembles also at the $SU(3)_F$ point \cite{Bruno:2016plf}. A determination using the gradient flow correlators with Wilson-Clover fermions calculated previously in \cite{Francis:2019muy} is given in yellow. The horizontal green band denotes the continuum extrapolated result of \cite{Bruno:2016plf}.

\section{Summary}

Through its open science philosophy OpenLat hopes to generate ensembles that  benefit the lattice QCD community. In our ongoing effort we continue to expand our repertoire of available ensembles, enhance our understanding and control over the simulations performed as well as increase the accessible parameter window for a better control of systematics in lattice observables. 

In this proceedings article we discussed three outcomes through OpenLat's research and ensembles: Firstly we gained more insights directly related to the generation process through starting to determine the distributions of $\lambda(\hat Q)$ and checking the regularity of the reweighting factor signs associated with this observable. We gained additional insights through continuing to tune and improve our set-ups. Secondly we reported on first successes in performing simulations at the physical point at $a=0.094~$fm lattice spacing in a large volume. Finally, we updated our determination of the renormalized pion decay constant as one of our initial observables.

In the near future we hope to expand our runs towards the second part of stage 2 and a set of ensembles at $m_\pi\simeq 200~$MeV, while continuing to push for the physical point at all lattice spacings available.
We remark that the procedures and results presented in this contribution are still preliminary and will be superseded by their final versions in the upcoming publication.

\section*{Acknowledgements}
OpenLat acknowledges support from the HPC computing centres hpc-qcd (CERN), HPE Apollo Hawk (HLRS) under grant stabwf/44185, NERSC (Cori and Perlmutter) through awards NP-ERCAP0024080, NP-ERCAP0020427, NP-ERCAP0017010, and NP-ERCAP0014740, Frontera (TACC), Piz Daint (CSCS), Occigen (CINES), Jean-Zay (IDRIS) and Ir\`ene-Joliot-Curie (TGCC) under projects 2023-A0140502271, (2020,2021,2022)-A0080511504, (2020,2021,2022)-A0080502271 by GENCI and PRACE project 2021250098. This work also used the DiRAC Extreme Scaling service at the University of Edinburgh, operated by the Edinburgh Parallel Computing Centre on behalf of the STFC DiRAC HPC Facility. DiRAC is part of the UK National e-Infrastructure.
AS acknowledges funding support under the National Science Foundation grant PHY-2209185. AF acknowledges support by the National Science and Technology Council of Taiwan under grant 111-2112-M-A49-018-MY2. We greatly acknowledge the leading role of Martin L\"uscher during the development and implementation of the SWF framework.

\bibliographystyle{JHEP}
\bibliography{references}


\end{document}